# Acoustic phonon dynamics in thin-films of the topological insulator $Bi_2Se_3$

Yuri D. Glinka,[1,2]* Sercan Babakiray,[1] Trent A. Johnson,[1] Mikel B. Holcomb,[1] and David Lederman[1]

[1]*Department of Physics and Astronomy, West Virginia University, Morgantown, WV 26506-6315, USA*
[2]*Institute of Physics, National Academy of Sciences of Ukraine, Kiev 03028, Ukraine*

Transient reflectivity traces measured for nanometer-sized films (6 – 40 nm) of the topological insulator $Bi_2Se_3$ revealed GHz-range oscillations driven within the relaxation of hot carriers photoexcited with ultrashort (~100 fs) laser pulses of 1.51 eV photon energy. These oscillations have been suggested to result from acoustic phonon dynamics, including coherent longitudinal acoustic phonons in the form of standing acoustic waves. An increase of oscillation frequency from ~35 to ~70 GHz with decreasing film thickness from 40 to 15 nm was attributed to the interplay between two different regimes employing traveling-acoustic-waves for films thicker than 40 nm and the film bulk acoustic wave resonator (FBAWR) modes for films thinner than 40 nm. The amplitude of oscillations decays rapidly for films below 15 nm thick when the indirect intersurface coupling in $Bi_2Se_3$ films switches the FBAWR regime to that of the Lamb wave excitation. The frequency range of coherent longitudinal acoustic phonons is in good agreement with elastic properties of $Bi_2Se_3$.

## I. INTRODUCTION

The study of topological insulators (TIs) (such as $Bi_2Se_3$, for example) is one of the most active areas in condensed matter physics due to unique properties of these materials. The combination of strong spin-orbit coupling and time-reversal symmetry in TIs allows for the existence of metallic two-dimensional (2D) Dirac surface states (SS), which have been established using angle-resolved photoemission spectroscopy (ARPES).[1,2] However, because surface effects can be easily masked by three-dimensional (3D) free carriers of the insulating medium (bandgap of bulk $Bi_2Se_3$ ~0.3 eV),[3-10] the control of purely Dirac-SS-governed properties of TIs remains a topic of interest. The problem can be partially solved by using the compensation doping method or a backgate voltage technique,[11,12] which efficiently deplete 3D carriers and hence switch electronic properties of the material to those related to 2D Dirac SS. The 3D carrier depletion effect can also be achieved in thin films of the TI $Bi_2Se_3$ due to indirect intersurface coupling, which is induced by the band structure distortion near the surface due to space-charge accumulation. Consequently, the depletion of 3D electrons across the film thickness occurs when the sum of depletion layer widths associated with each surface of the film exceeds the film thickness.[13,14] This kind of indirect intersurface coupling usually occurs over a length scale (≤15 nm) much greater than the critical thickness (6 nm) of direct coupling[15] and has been recognized using ultrafast photoexcited carrier relaxation dynamics[13] and Raman spectroscopy.[14] The indirect intersurface coupling effect can also be studied using other surface-sensitive techniques. One of such techniques employing optical second harmonic generation (SHG) is known to be sensitive to the depletion electric field and therefore seems to be suitable for this purpose, especially if it is realized in the ultrafast pump-probe configuration.[16-18] Another technique involves the propagation of surface acoustic waves (SAW - Rayleigh waves), which has successfully been used to study 2D electron gas in GaAs heterostructures[19,20] and magneto-conductance in Dirac cone quasiparticles of graphene.[21] The excitation of Rayleigh waves in $Bi_2Te_3$ and $Bi_2Se_3$ single crystals has been predicted theoretically[22-24] and therefore the application of SAW seems very promising for studying Dirac SS in TIs. However, in thin films the dynamics of coherent acoustic waves are known to switch to those related to the film bulk acoustic wave resonator (FBAWR) modes.[25] Because FBAWR modes are completely governed by the resonator width (film thickness), the study of acoustic waves in thin films of the TI $Bi_2Se_3$ is expected to provide knowledge for a more thorough understanding of indirect intersurface coupling and its effect on carrier transport phenomena.

It should be noted that previous studies of ultrafast carrier dynamics in $Bi_2Se_3$ single crystals revealed one cycle oscillations observed in the transient reflectivity (TR) traces, which were attributed to coherent acoustic phonons.[26,27] Furthermore, the relative independence of resistivity of Dirac SS in thin exfoliated $Bi_2Se_3$ from 2D carrier density has been suggested to result from intrinsic electron-acoustic-phonon scattering that limits charge carrier mobility at room temperature.[28] These findings suggest that coherent acoustic phonon dynamics in $Bi_2Se_3$ thin films have not just a pure scientific interest but also would be of great importance for potential applications of TIs in novel nanoscale devices.

In this paper, we report on thickness-dependent GHz-range oscillations in $Bi_2Se_3$ thin-films (15 to 40 nm thick) observed in pump-probe TR traces. The oscillations were generated within the relaxation of hot carriers photoexcited with ultrashort (~100 fs) laser pulses of 1.51 eV photon energy. Our findings suggest about the acoustic wave nature of these oscillations and the Brillouin backscattering mechanism for their observation. We experimentally identified the interplay between two regimes, for which the traveling-acoustic-waves and the FBAWR modes predominantly contribute into the acoustic wave dynamics with decreasing film thickness. Furthermore, we found that the amplitude of oscillations decays rapidly for the thinnest films (<15 nm) when indirect intersurface coupling switches from the FBAWR regime to that of the Lamb wave excitation.

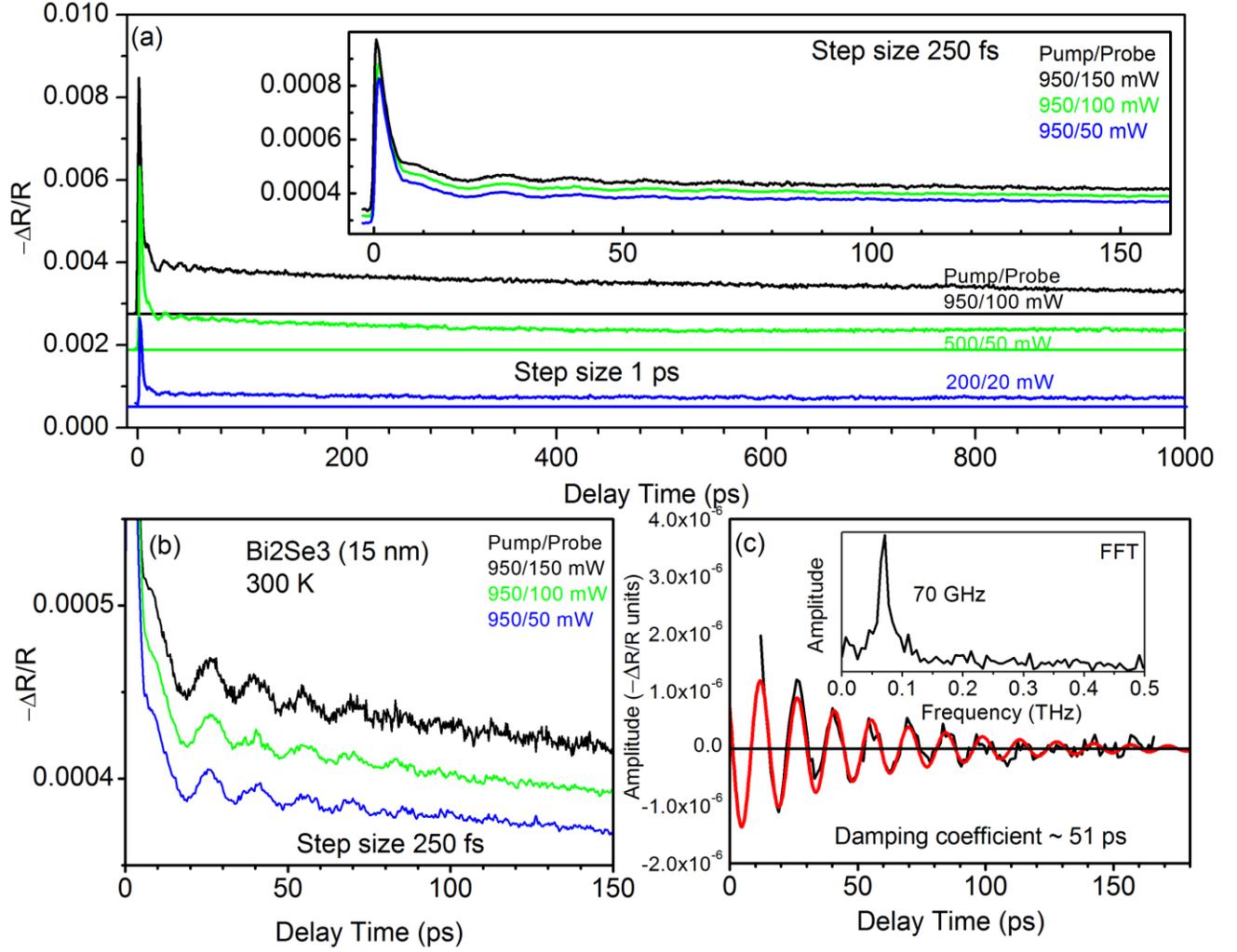

FIG. 1. (a) TR traces for the 15 nm thick $Bi_2Se_3$ film measured with 1-ps step size and with varying pump-probe average power indicated by the corresponding colors. The same color horizontal lines present individual baselines for each of the traces. Inset shows TR traces measured at the same experimental conditions but with a step size of 250 fs and with varying probe average power indicated by the corresponding colors. (b) TR traces shown in the inset of (a) but re-scaled for better observation of oscillations. (c) The extracted oscillatory part of the TR traces shown in (b) (in black) and a best fit by a damped sine function with a damping coefficient of ~ 51 ps (in red). Inset shows the FFT of the oscillatory part with the center frequency at ~70 GHz.

## II. SAMPLE AND EXPERIMENTAL SETUP

Experiments were performed on $Bi_2Se_3$ thin-films of 6, 8, 10, 12, 15, 20, 25, 30, 35, and 40 quintuple layers (QL ~ 1 nm) thick using a Ti:Sapphire laser with pulse duration 100 fs, center photon energy 1.51 eV (820 nm) and repetition rate 80 MHz. The films were grown on 0.5 mm $Al_2O_3(0001)$ substrates by molecular beam epitaxy, with a 10 nm thick $MgF_2$ capping layer to protect against oxidation. The growth procedure and the optical experimental setup were similar to those described previously.[13,29] The free-carrier densities in the films were measured using the Hall effect,[30] and were found to be typical for as-grown $Bi_2Se_3$ films and single crystals (0.5 – $3.5\times10^{19}$ $cm^{-3}$). No film damage was observed for the laser powers used in this study, eliminating the heating effects from our consideration.

## III. EXPERIMENTAL RESULTS AND DISCUSSION

Figure 1 shows an example of TR traces measured for the 15 nm thick $Bi_2Se_3$ film at different pump-probe average powers. The same measurements were performed for all available samples of various $Bi_2Se_3$ film thicknesses. The general behavior of TR traces is similar to those reported previously,[13] except for a weak oscillatory part with frequency of ~70 GHz (for the 15 nm thick film) which is superimposed over the long-lived decay component.[13,29] The oscillatory part of TR signals weakly depends on the pump and probe powers, indicating that oscillations are nearly independent of photoexcited carrier densities [Fig. 1(a) and (b)]. To extract the oscillatory part, the TR traces were fitted by a multi-exponential rise-decay function as described previously [Fig. 1(c)].[13,29] Subsequently, a fast Fourier transformation (FFT) was applied to obtain the oscillation frequency [Fig. 1(c) Inset]. Finally, we used a damped sine function to fit the oscillatory part and to obtain the corresponding amplitudes and damping coefficients.

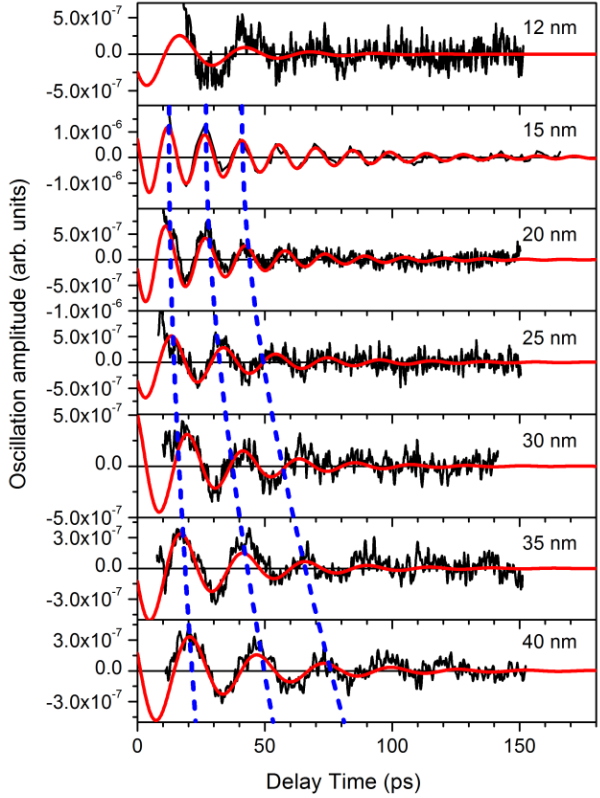

FIG. 2. The extracted oscillatory part of the TR traces shown as a function of $Bi_2Se_3$ film thickness indicated. The best fit by a damped sine function is shown in red. The blue dashed curves present the guide for eyes of the positions of the corresponding peaks, indicating an increase of the frequency of oscillations with decreasing film thickness in the range from 40 nm to 15 nm.

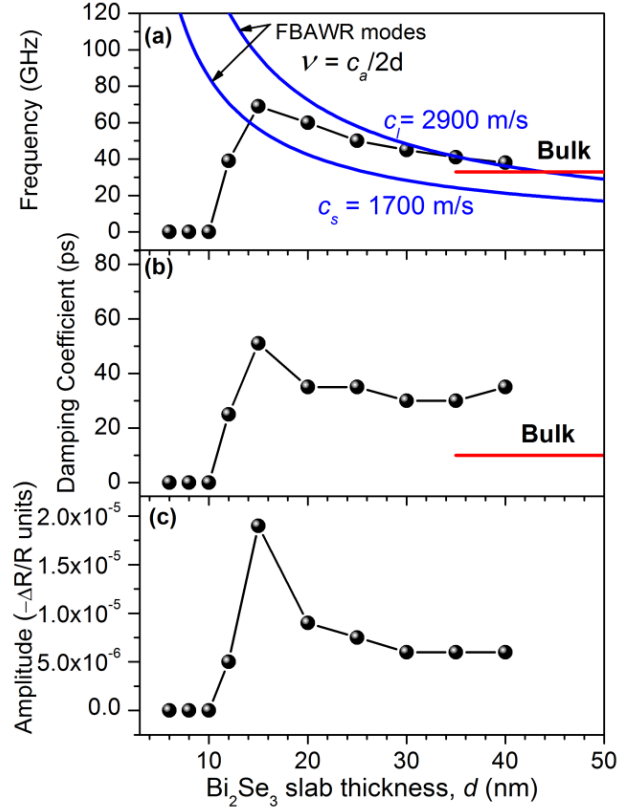

FIG. 3. $Bi_2Se_3$ film thickness dependences of the frequency (a), damping coefficient (b), and amplitude (c) of the oscillatory part of the measured TR traces. Blue curves in (a) show the predicted evolution of FBAWR fundamental mode frequency ($\nu = c_a/2d$) with $Bi_2Se_3$ film thickness ($d$) for longitudinal ($l$) and shear ($s$) acoustic waves of $c_a = c_l = 2900$ m/s and $c_a = c_s = 1700$ m/s speed of sound,[24] respectively. Red horizontal lines in (a) and (b) present the corresponding bulk parameters.[26,27]

Figure 2 shows that the frequency, damping coefficient, and amplitude of oscillations gradually increase with decreasing film thickness in the range from 40 nm to 15 nm. These dynamics is followed by a complete suppression of oscillations for the thinner films (6 to 12 nm thick). Figure 3 summarizes the oscillatory dynamics of TR traces as a function of the $Bi_2Se_3$ film thickness and as compared to the $Bi_2Se_3$ single crystal parameters.[26,27]

For thicker films (40 nm) the frequency of oscillations (38 GHz) was slightly above that observed for $Bi_2Se_3$ single crystals (33 GHz) and assigned to the coherent acoustic phonons.[26,27] In contrast, the damping coefficient obtained for the 40 nm thick film is more than two times larger than that for $Bi_2Se_3$ single crystals. This behavior suggests that even in the thicker films used in this study, the film boundaries strongly affect acoustic phonon dynamics. In general, the excitation of acoustic phonons by ultrashort laser pulses is expected to occur indirectly through the anharmonic three-phonon (Klemens/Ridley) decay of longitudinal optical (LO) phonons involved into the hot carrier relaxation process.[13,29,31] This statement is based on the fact that the electron-phonon coupling strength in $Bi_2Se_3$ deduced from inelastic helium-atom scattering (~0.43)[32] points to the extremely strong electron-phonon coupling regime in the bulk, which can be associated with electron-polar-phonon (Fröhlich) coupling in a similar way as that in the wide bandgap materials such as GaN (~0.49), where the Klemens/Ridley process governs the relaxation dynamics.[31,33] Alternatively, the exceptionally weak electron-phonon coupling regime in $Bi_2Se_3$ with a constant of ~0.08 deduced from ARPES points to electron-phonon coupling that occurs in Dirac SS,[34] since the latter constant is similar to typical values of good conductors such as noble metals (~ 0.1),[35] where other sources of electron-phonon coupling associated with deformation potential or thermoelastic scattering can dominate the intrasubband relaxation.[36] Because the carriers photoexcited in $Bi_2Se_3$ films with 1.51 eV photons mainly relax through the bulk,[13,29] our assumption seems reasonable. Also, we note that the frequency of 38 GHz (0.157 meV) observed for the 40 nm thick film closely matches the range of Rayleigh waves theoretically predicted for $Bi_2Te_3$ single crystals.[23] The reference to $Bi_2Te_3$ seems to be suitable since $Bi_2Te_3$ and $Bi_2Se_3$ have similar elastic properties.[23,24] However, despite this coincidence, the Rayleigh wave (SAW) model would be unable to explain the thickness dependence of the observed oscillations.

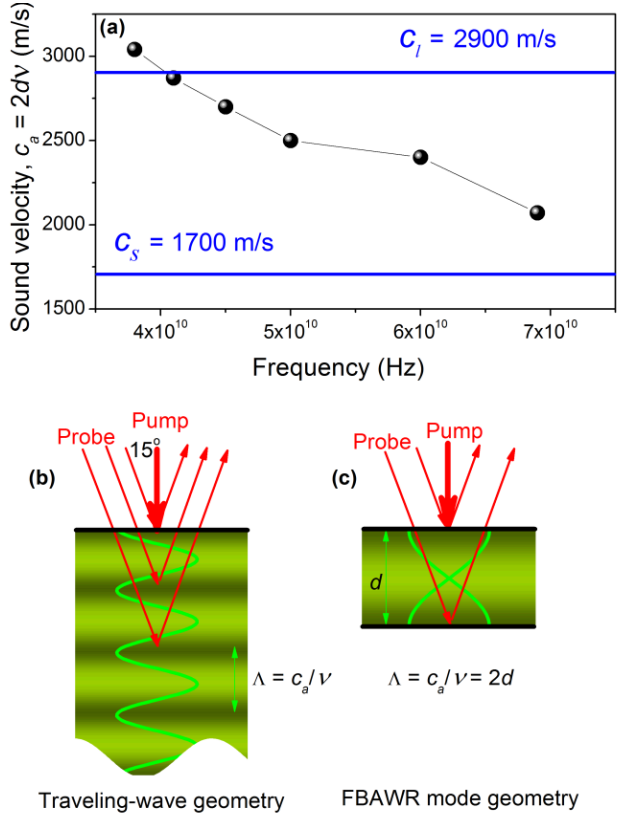

FIG. 4. (a) Experimentally obtained sound velocity versus the frequency of oscillatory part of the measured TR traces. (b) and (c) Schematic illustrations of the travelling longitudinal/shear waves and FBAWR modes excited by the laser pump and probed through the Brillouin backscattering mechanism, respectively.

Consequently, we associate the oscillatory behavior of TR traces for thicker films (as well as for single crystals) of $Bi_2Se_3$ with the excitation of the traveling longitudinal acoustic wave viewed through the Brillouin backscattering mechanism.[37-41] It is well known that owing to the significant excess of the optical decay rate over the acoustic mode decay rate, the Brillouin scattering traveling-wave geometry acts as an acoustic phonon amplifier.[39,40] This model can be convincingly applied since the light penetration depth in $Bi_2Se_3$ thin-films is known to be always shorter than the film thickness for thicknesses in the range of 15 - 40 nm.[29] If the traveling wave is affected by boundaries, the acoustic phonon amplification manifests itself as an enhancement of the resonant fundamental frequency mode of the nanocavity, which acts as an acoustic resonator and is determined exclusively by the film thickness.[37,39] Alternatively, the coupled nanocavity structures in phononic crystals are known to reveal an acoustic frequency comb which originates from a series of nanocavities of different thicknesses.[38]

As a result, the increase of oscillation frequency with decreasing film thickness [Fig. 3(a)] points to a crossover between the two regimes for which traveling-acoustic-waves and FBAWR modes predominantly contribute to the Brillouin backscattering process. The resulting FBAWR modes excited in thicker films (35 – 40 nm) revealed the characteristic dependence of the fundamental resonant frequency as a function of thickness ($d$) of $\nu = c_a/2d$, where $c_a$ is the sound velocity for the longitudinal/transverse (shear) acoustic waves.[25] To identify the FBAWR efficiency, the sound reflectivity coefficients[17] for the $Al_2O_3$/$Bi_2Se_3$ and $Bi_2Se_3$/$MgF_2$ interfaces can be estimated as $R_1 = |(z_{Al2O3} - z_{Bi2Se3})/(z_{Al2O3} + z_{Bi2Se3})| = 0.182$ and $R_2 = |(z_{MgF2} - z_{Bi2Se3})/(z_{MgF2} + z_{Bi2Se3})| = 0.131$, respectively, where acoustic impedances for $Al_2O_3$ ($z_{Al2O3} = 31.48 \times 10^6$ kg/sm$^2$), $Bi_2Se_3$ ($z_{Bi2Se3} = 21.78 \times 10^6$ kg/sm$^2$), and $MgF_2$ ($z_{MgF2} = 16.74 \times 10^6$ kg/sm$^2$) have been taken into account.[42] The small reflectivity coefficients suggest the existence of significant losses in FBAWR and hence agree well with a short life-time of the FBAWR modes observed [Fig. 3(b)].

Taking into account the theoretically determined sound velocities of longitudinal and shear waves in $Bi_2Se_3$ as $c_a = c_l = 2900$ m/s and $c_a = c_s = 1700$ m/s, respectively,[24] the FBAWR fundamental mode frequencies can be plotted as a function of film thickness [Fig. 3(a)]. The resulting frequency variation is in good agreement with experimentally observed frequencies and hence suggests a good consistency with elastic properties of $Bi_2Se_3$. However, the frequencies observed for thicker films match closely the FBAWR frequencies for the longitudinal acoustic wave velocity, whereas the acoustic waves begin demonstrating a shear wave behavior with decreasing film thickness. These dynamics are shown more clearly in Figure 4(a), where the sound velocity ($c_a = 2d\nu$) is plotted against the FBAWR mode frequency ($\nu$). We associate this tendency with coupling between longitudinal and shear FBAWR fundamental modes, which increases gradually with decreasing film thickness. This behavior implies that boundary conditions lead to coupling between the longitudinal and transverse potentials so that FBAWR mode has both longitudinal and transverse character.[43] Also, our findings for $Bi_2Se_3$ thin films suggest that the FBAWR mode regime, for which the sound velocity is usually assumed to be constant,[25] is significantly modified by the dependence of the sound velocity on film thickness.

According to the Brillouin scattering mechanism,[41] the coherent acoustic phonon replicas in TR traces result from the constructive interference of the multiple light beams scattered by the plane acoustic wave, generating a periodic compressive lattice strains which modulate the local dielectric constant (and hence the refractive index) of $Bi_2Se_3$. Because of the large discrepancy between the sound velocity (~$10^3$ m/s) and the velocity of the probing light (~$10^8$ m/s), the dielectric inhomogeneities can be regarded as a quasi-static lattice on which the photons of the probing light are scattered. Consequently, $Bi_2Se_3$ can be treated as a periodic multilayer stack with periodicity governed by the wavelength of the acoustic wave excited ($\Lambda$). The reflected intensity reaches its maximum when $\Lambda = \lambda / 2n_r \sin\left[(180° - 2\varphi)/2\right]$,[41] where the refractive index of $Bi_2Se_3$ $n_r = 5.5$,[44] the probe beam incidence angle used $\varphi = 15°$, and the wavelength of laser light $\lambda = 820$ nm should be taking into account. Consequently, the obtained value of Λ = 77 nm corresponds to the frequency of oscillation for longitudinal acoustic wave ($\nu = c_l/\Lambda$) of 37.7 GHz, which closely matches that observed for the 40 nm thick film (38 GHz). Furthermore, assuming the FBAWR mode regime (Λ = 2$d$), the obtained value of Λ is relevant for the thicker Bi$_2$Se$_3$ films ($d$ = 35 - 40 nm). These estimates confirm that the acoustic waves in thicker films of Bi$_2$Se$_3$ as well as in single crystals travel with the longitudinal wave velocity [Fig. 3(a) and 4(b)]. One can estimate that by changing the probe beam incidence angle to that of widely used in pump-probe measurements ($\varphi$ = 45°), the acoustic wavelength changes to Λ = 105 nm, which corresponds to the traveling-acoustic-wave frequency of 28 GHz. The latter value still remains close to that experimentally observed for Bi$_2$Se$_3$ single crystals (33 GHz).[26,27] When the film thickness decreases in the range from 40 nm to 15 nm, the sound velocity decreases due to the additional shear lattice strains involved. Nevertheless, the FBAWR modes still dominate the acoustic wave dynamics and the frequency of oscillations increases with decreasing film thickness according to the same $\nu = c_a/2d$ law, but also taking into account the decrease of the sound velocity [Fig. 4(c)]. This behavior explains the deviation of experimentally observed frequencies with decreasing film thickness from those predicted for purely longitudinal acoustic waves [Fig. 3(a)].

The suppression of oscillations for films below 15 nm thick points to another regime reached in acoustic wave dynamics, at which indirect intersurface coupling completely destroys the FBAWR mode behavior as a consequence of the loss of the flexural rigidity of the film initiated by the progressive shear lattice strains. The indirect intersurface coupling leads to the joint out-of-phase/in-phase (symmetric/asymmetric with respect to the mid-plane of the film) motions of both surfaces of the films, which corresponds to dominant shear lattice strains in the film and is associated with the Lamb wave excitation.[41] Because longitudinal acoustic waves involve changes in the volume of the Bi$_2$Se$_3$ films, i.e., dilatation/compression of a local volume element (and hence in the refractive index) while the shear waves cause no volume change, the oscillations in TR traces become suppressed with decreasing film thickness in the range below 15 nm (Fig. 3). This behavior implies that the pump-probe TR technique in the configuration used in the current study is not suitable for monitoring shear Lamb waves in thin-films of the TI Bi$_2$Se$_3$ with thickness below 15 nm. We also note that our theoretical prediction implemented in Figure 3(a) for shear acoustic velocity allows the frequency range as high as ~200 GHz to be reached for Bi$_2$Se$_3$ films where the direct intersurface coupling occurs (≤6 nm).

## IV. CONCLUSIONS

We have provided experimental evidence for acoustic wave excitations in nanometer-sized films of the TI Bi$_2$Se$_3$, which were excited within the relaxation of hot carriers photoexcited with ultrashort (~100 fs) laser pulses of 1.51 eV photon energy. The GHz range oscillations were viewed through the Brillouin backscattering mechanism and originate from the longitudinal traveling-acoustic-waves and FBAWR modes, which predominantly contribute to TR for ≥40 nm films and those in the thickness range of 40 - 15 nm, respectively. The rapid amplitude decay of oscillations for the thinnest films (<15 nm) has been attributed to the indirect intersurface coupling, which switches the FBAWR mode dynamics to those of the Lamb wave excitation.

## ACKNOWLEDGMENTS

This work was supported by a Research Challenge Grant from the West Virginia Higher Education Policy Commission (HEPC.dsr.12.29). Some of the work was performed using the West Virginia University Shared Research Facilities.

[1] S. Murakami, New J. Phys. **9**, 356 (2007).
[2] M. Z. Hasan, C. L. Kane, Rev. Mod. Phys., **82**, 3045 (2010).
[3] D. Hsieh, Y. Xia, D. Qian, L. Wray, J. H. Dil, F. Meier, J. Osterwalder, L. Patthey, J. G. Checkelsky, N. P. Ong, A. V. Fedorov, H. Lin, A. Bansil, D. Grauer, Y. S. Hor, R. J. Cava, and M. Z. Hasan, Nature (London) **460**, 1101 (2009).
[4] Y. L. Chen, J.-H. Chu, J. G. Analytis, Z. K. Liu, K. Igarashi, H.-H. Kuo, X. L. Qi, S. K. Mo, R. G. Moore, D. H. Lu, M. Hashimoto, T. Sasagawa, S. C.Zhang, I. R. Fisher, Z. Hussain, and Z. X. Shen, Science **329**, 659 (2010).
[5] L. A. Wray, S.-Y. Xu, Y. Xia, Y. S. Hor, D. Qian, A. V. Fedorov, H. Lin, A. Bansil, R. J. Cava, and M. Z. Hasan, Nat. Phys. **6**, 855 (2010).
[6] J. A. Sobota, S. Yang, J. G. Analytis, Y. L. Chen, I. R. Fisher, P. S. Kirchmann, and Z.-X. Shen, Phys. Rev. Lett. **108**, 117403 (2012).
[7] Y. H. Wang, D. Hsieh, E.J. Sie, H. Steinberg, D.R. Gardner, Y.S. Lee, P. Jarillo-Herrero, and N. Gedik, Phys. Rev. Lett. **109**, 127401 (2012).
[8] Y. M. Hajlaoui, E. Papalazarou, J. Mauchain, L. Perfetti, A. Taleb-Ibrahimi, F. Navarin, M. Monteverde, P. Auban-Senzier, C.R. Pasquier, N. Moisan, D. Boschetto, M. Neupane, M.Z. Hasan, T. Durakiewicz, Z. Jiang, Y. Xu, I. Miotkowski, Y.P. Chen, S. Jia, H.W. Ji, R.J. Cava and M. Marsi, Nature Comm. **5**, 3003 (2014).
[9] Y. H. Wang, H. Steinberg, P. Jarillo-Herrero, and N. Gedik, Science **342**, 453 (2013).
[10] J. A. Sobota, S.-L. Yang, A. F. Kemper, J. J. Lee, F. T. Schmitt, W. Li, R. G. Moore, J. G. Analytis, I. R. Fisher, P. S. Kirchmann, T. P. Devereaux, and Z.-X. Shen, Phys. Rev. Lett. **111**, 136802 (2013).
[11] M. Brahlek, N. Koirala, M. Salehi, N. Bansa, and S. Oh, Phys. Rev. Lett. **113**, 026801 (2014).
[12] S. Cho, B. Dellabetta, A. Yang, J. Schneeloch, Z. Xu, T. Valla, G. Gu, M. J. Gilbert, and N. Mason, Nat. Comm. **4**, 1689 (2013).
[13] Y. D. Glinka, S. Babakiray, T. A. Johnson, M. B. Holcomb, and D. Lederman, Appl. Phys. Lett. **105**, 171905 (2014).
[14] Y. D. Glinka, S. Babakiray, T. A. Johnson, and D. Lederman, J. Phys.: Condens. Matter **27**, 052203 (2015).
[15] Y. Zhang, K. He, C.-Z. Chang, C.-L. Song, L.-L. Wang, X. Chen, J.-F. Jia, Z. Fang, X. Dai, W.-Y. Shan, S.-Q. Shen, Q. Niu, X.-L. Qi, S.-C. Zhang, X.-C. Ma, and Q.-K. Xue, Nature Phys. **6**, 584 (2010).

[16] T. A. Germer, K. W. Kolasinski, J. C. Stephenson, and L. J. Richter, Phys. Rev. B **55**, 10694 (1997).

[17] Y. D. Glinka, N. H. Tolk, and J. K. Furdyna, Phys. Rev. B **84**, 153304 (2011).

[18] D. Hsieh, F. Mahmood, J. W. McIver, D. R. Gardner, Y. S. Lee, and N. Gedik, Phys. Rev. Lett. **107**, 077401 (2011).

[19] A. Wixforth, J. P. Kotthaus, and G. Weimann, Phys. Rev. Lett. **56**, 2104 (1986).

[20] R. L. Willett, M. A. Paalanen, R. R. Ruel, K. W. West, L. N. Pfeiffer, and D. J. Bishop, Phys. Rev. Lett. **65**, 112 (1990).

[21] P. Thalmeier, B. D´ora, and K. Ziegler, Phys. Rev. B **81**, 041409(R) (2010).

[22] P. Thalmeier, Phys. Rev. B **83**, 125314 (2011).

[23] S. Giraud and R. Egger, Phys. Rev. B **83**, 245322 (2011).

[24] S. Giraud, A. Kundu, and R. Egger, Phys. Rev. B **85**, 035441 (2012).

[25] J. Berge, M. Norling, A. Vorobiev, and S. Gevorgian, J. Appl. Phys. **103**, 064508 (2008).

[26] J. Qi, X. Chen, W. Yu, P. Cadden-Zimansky, D. Smirnov, N. H. Tolk, I. Miotkowski, H. Cao, Y. P. Chen, Y. Wu, S. Qiao, and Z. Jiang, Appl. Phys. Lett. **97**, 182102 (2010).

[27] N. Kumar, B. A. Ruzicka, N. P. Butch, P. Syers, K. Kirshenbaum, J. Paglione, and H. Zhao, Phys. Rev. B **83**, 235306 (2011).

[28] D. Kim, Q. Li, P. Syers, N. P. Butch, J. Paglione, S. Das Sarma, and M. S. Fuhrer, Phys. Rev. Lett. **109**, 166801 (2012).

[29] Y. D. Glinka, S. Babakiray, T. A. Johnson, A. D. Bristow, M. B. Holcomb, and D. Lederman, Appl. Phys. Lett. **103**, 151903 (2013).

[30] P. Tabor, C. Keenan, S. Urazdhin, and D. Lederman, Appl. Phys. Lett. **99**, 013111 (2011).

[31] G. P. Srivastava, J. Phys.: Condens. Matter **21**, 174205 (2009).

[32] X. Zhu, L. Santos, C. Howard, R. Sankar, F. C. Chou, C. Chamon, and M. El-Batanouny, Phys. Rev. Lett. **108**, 185501 (2012).

[33] M. A. Reshchikov and H. Morkoç, J. Appl. Phys. **97**, 061301 (2005).

[34] Z.-H. Pan, A.V. Fedorov, D. Gardner, Y. S. Lee, S. Chu, and T. Valla, Phys. Rev. Lett. **108**, 187001 (2012).

[35] R. C. Hatch, M. Bianchi, D. Guan, S. Bao, J. Mi, B. B. Iversen, L. Nilsson, L. Hornekær, and P. Hofmann, Phys. Rev. B **83**, 241303 (2011).

[36] V. I. Pipa, N. Z. Vagidov, V. V. Mitin, and M. Stroscio, Phys. Rev. B **64**, 235322 (2001).

[37] R. P. Beardsley, A.V. Akimov, M. Henini, and A. J. Kent, Phys. Rev. Lett. **104**, 085501 (2010).

[38] N. D. Lanzillotti-Kimura, A. Fainstein, B. Perrin, B. Jusserand, O. Mauguin, L. Largeau, and A. Lema1tre, Phys. Rev. Lett. **104**, 197402 (2010).

[39] I. S. Grudinin, H. Lee, O. Painter, and K. J. Vahala, Phys. Rev. Lett. **104**, 083901 (2010).

[40] R.Y. Chiao, C. H. Townes, and B. P. Stoicheff, Phys. Rev. Lett. **12**, 592 (1964).

[41] T. Still, *High Frequency Acoustics in Colloid-Based Meso- and Nanostructures by Spontaneous Brillouin Light Scattering* (Springer, Berlin, 2010).

[42] H. Wang, Y. Xu, M. Shimono, Y. Tanaka, and M. Yamazaki, Mat. Transactions, **48**, 2349 (2007).

[43] A. R. Duggal, J. A. Rogers, and K. A. Nelson, J. Appl. Phys. **72**, 2823 (1992).

[44] J. W. McIver, D. Hsieh, S. G. Drapcho, D. H. Torchinsky, D. R. Gardner, Y. S. Lee, and N. Gedik, Phys. Rev. B **86**, 035327 (2012).